# Identification of Shared Genetic Biomarkers to Discover Candidate Drugs for Cervical and Endometrial Cancer by Using the Integrated Bioinformatics Approaches

Md. Selim Reza[1], Mst. Ayesha Siddika[3], Md. Tofazzal Hossain[4], Md. Ashad Alam[5], Md. Nurul Haque Mollah[2]

[1] Tulane Center of Biomedical Informatics and Genomics, Tulane University, New Orleans, LA 70112, USA

[2] Bioinformatics Lab, Department of Statistics, University of Rajshahi, Rajshahi 6205, Bangladesh

[3] Microbiology Lab., Department of Veterinary and Animal Sciences, University of Rajshahi, Rajshahi-6205, Bangladesh

[4] Department of Statistics, Bangabandhu Sheikh Mujibur Rahaman Science and Technology University, Gopalganj 8100, Bangladesh,

[5] Ochsner Center for Outcomes Research, Ochsner Research, Ochsner Clinic Foundation, New Orleans, LA 70121, USA

## Abstract

Cervical (CC) and endometrial cancers (EC) are two common types of gynecological tumors that threaten the health of females worldwide. Since their underlying mechanisms and associations remain unclear, computational bioinformatics analysis is required. In the present study, bioinformatics methods were used to screen for key candidate genes, their functions and pathways, and drug agents associated with CC and EC, aiming to reveal the possible molecular-level mechanisms. Four publicly available microarray datasets of CC and EC from the Gene Expression Omnibus database were downloaded, and 72 differentially expressed genes (DEGs) were selected through integrated analysis. Then, we performed the protein-protein interaction (PPI) analysis and identified 9 shared genetic biomarkers (SGBs). The GO functional and KEGG pathway enrichment analyses of these SGBs revealed some important functions and signaling pathways significantly associated with CC and EC. The interaction network analysis identified four transcription factors (TFs) and two miRNAs as key transcriptional and post-transcriptional regulators of SGBs. The expression of the AURKA, TOP2A, and UBE2C genes was higher in CC and EC tissues than in normal samples, and this gene expression was linked to disease progression. Furthermore, we performed docking analysis between 9 SGBs-based proteins and 145 meta-drugs, and identified the top-ranked 10 drugs as candidate drugs. Finally, we investigated the binding stability of the top-ranked three drugs (Sorafenib, Paclitaxel, Sunitinib) using 100 ns MD-based MM-PBSA simulations with UBE2C, AURKA, and TOP2A proteins, and observed their stable performance. Therefore, the proposed drugs might play a vital role in the treatment against CC and EC.

**Keywords:** cervical cancer, endometrial carcinoma, bioinformatics, Docking, MD-Simulations

## 1. Introduction

Gynecological tumors are a specific kind of malignant tumor that are developed in the female reproductive system and poses a major threat to the patient's life. Among the various gynecological malignancies, endometrial carcinoma (EC) and cervical cancer (CC) are the top two most prevalent tumors of the female reproductive system, alongside ovarian cancer [1,2]. Gynecological cancer continues to place a considerable burden on global healthcare systems, despite overall declines in incidence and death rates driven by improved understanding of the disease [1]. Early diagnosis and therapy are crucial for improving patient outcomes, but they also require a better understanding of the disease's molecular pathophysiology, as well as the development of suitable biomarkers and therapeutic targets. Previous research has shown that the prevalence of CC and human papillomavirus (HPV) infection are closely related [3]. According to pathogenetic theories, EC is also divided into two categories based on estrogen dependence [4]. The precise pathogenesis of these two forms of cancer is still unknown, despite several earlier studies on their etiology.

There are certain pathological and etiological associations between CC and EC, as both are squamous cell cancers and both are associated with HPV infection [5,6] Unlike CC, EC is connected with sex hormones, paralleling other aggressive cancers in females, including ovarian and breast cancer [4]. Moreover, pathology plays a major role in the clinical diagnosis of these two cancer types [7]. Precise biomarkers remain unknown in the early stages of CC and EC. The paramesonephric ducts, which give rise to the entire female reproductive tract and evolve into various organs through a complex regulatory mechanism, are considered to be the embryological source of both cervical and endometrial tissues [8]. For this reason, despite the differences between CC and EC, it has been postulated that these two types of gynecological tumor have a common mechanism, and that certain marker molecules may be shared in their carcinogenesis and development. Therefore, in-silico studies might enhance understanding of these two categories of cancers.

The identification of genetic biomarkers could act as a catalyst for the development of future, more effective therapy strategies [9]. Genetic biomarkers may enhance the effectiveness of CC and EC diagnostic and therapeutic methods, according to their prospective advantages. Integrated bioinformatics approaches are now broadly used in numerous biological research fields. Recent progress in sequencing techniques has enabled researchers to make critical discoveries in computational biology and molecular treatments [10]. Protein-Protein Interactions (PPI) were previously used to detect potential genetic biomarkers responsible for the infection, and a co-expression network was employed to validate the listed genetic biomarkers using a heatmap based on their co-regulation scores [11,12] Protein 3-dimensional (3D) structures are important in fields like evolutionary biology and biotechnology, such as protein function prediction and drug design [13].

However, de-novo drug development is a difficult, time-consuming, and costly endeavor. The main challenges are to identify CC and EC disease-causing drug target receptors and drug agents that can alleviate disease severity by interacting with the target proteins. As opposed to developing a new drug, repurposing existing drugs for certain conditions might save time and money. To date, numerous studies have proposed different sets of hub/studied-genes to explore molecular mechanisms and pathogenetic processes in CC and EC [14,15]. Some of these studies have also suggested their hub-genes or studied-gene guided candidate drugs for the treatment of CC and EC [3,16]. By reviewing the literature, we observed that their suggested CC-causing or EC-causing hub-/studied-gene sets, as well as candidate drugs, are not consistent across different published articles. On the other hand, so far, none of these studies investigated the performance of their suggested drugs against the other published hub-genes mediated target proteins. Obviously, more representative hub-genes are required to explore more effective candidate drugs against CC or EC. Therefore, in this study, our main objectives are to (i) computationally identify shared genetic biomarkers (SGBs) target receptors between CC and EC, highlighting their functions, pathways, and regulatory factors, (ii) explore proposed shared genomic biomarkers-guided candidate drugs for the treatment against CC and EC, and (iii) in-silico validation of the proposed drugs against the state-of-the-art alternative top-ranked independent receptors proposed by others. The workflow of this study is demonstrated in Figure 1.

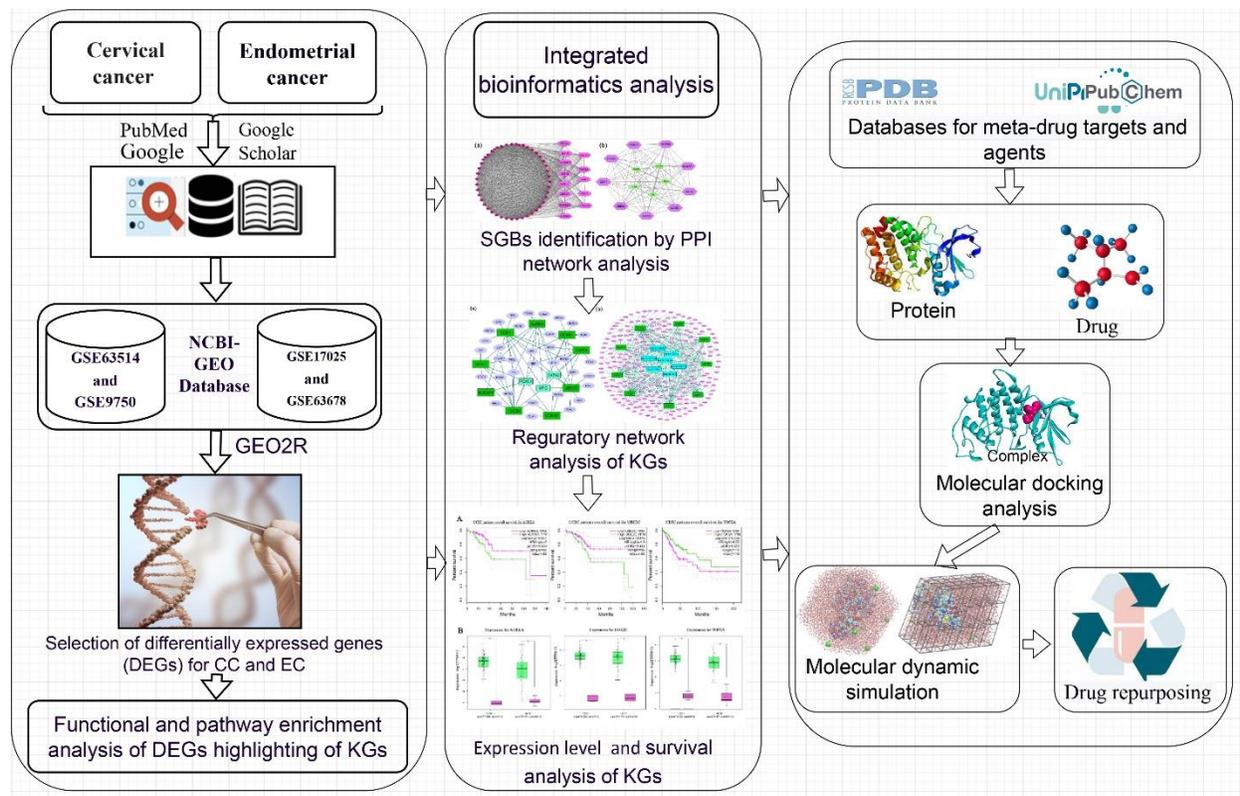

**Figure 1.** The pipeline of this study.

## 2. Materials and Methods

To explore the sharing CC and EC-causing genetic biomarkers highlighting their functions, pathways, regulatory factors, we utilized the statistical r-package LIMMA, online databases STRING, GO, KEGG, and DisGeNET, and online tools NetworkAnalyst, GEPIA2, and Cytoscape. To explore genetic biomarker-guided repurposable drugs, we performed molecular docking analysis by using offline tools (Discovery Studio Visualizer, Autodock Vina, PyMol, and Yasara). The detailed procedure is discussed below.

2.1. Data Sources and Descriptions

We used both original data and meta-data to reach the goal of this study as described below.

2.2. Microarray Data Information

We collected microarray profiles for cervical cancer (CC) and endometrial carcinoma (EC) disease from the National Center for Biotechnology Information (NCBI) Gene Expression Omnibus (GEO) website (http://www.ncbi.nlm.nih.gov/geo/)[17]. We used four publicly available microarray profiles, GSE63514 [18] and

GSE9750 [19] for cervical cancer (CC), and GSE17025 [20] and GSE63678 [21] for endometrial carcinoma (EC) to identify the shared DEGs that are associated with tumor grade. The microarray dataset of GSE63514 was based on the GPL570 Platforms ((HG-U133_Plus_2) Affymetrix Human Genome U133 Plus 2.0 Array), which included 28 CC tissues and 24 normal cervical tissues. The GSE9750 dataset used the GPL96 Affymetrix Human Genome U133A Array platform and included 33 CC tissue samples that were primarily marked by HPV16 or HPV18 and 21 normal samples. The datasets GSE17025 was based on the Affymetrix Human Genome U133 Plus 2.0 Array, which consists of 79 EC samples and 12 normal samples. The EC patient's dataset GSE63678 was generated using Affymetrix Human Genome U133A 2.0 Array technology with 5 normal tissues and 7 EC tissues; analysis was performed using platform GPL571.

2.3. Collection of studied-DEGs for Cross-Validation with the Proposed Drugs

To select the top-ranked hub genes (independent meta-receptors) associated with CC and EC disease, we reviewed 66 published articles and selected the top-ranked 7 target proteins as the independent meta-receptors (see Table S1).

2.4. Collection of Meta-Drug Agents for Exploring Candidate Drugs

We collected host-transcriptome-guided 145 meta-drug agents by the literature review of CC and EC disease (see Table S2) for exploring candidate drugs. Thus, we considered 145 drug agents to explore candidate drugs by molecular docking with our proposed receptors.

2.5. Differential expression analysis for CC and EC Patients

The identification of common differential expression genes (DEGs) was a key step of this study. To identify common-DEGs, we identified DEGs for each of GSE63514 and GSE9750, GSE17025 and GSE63678 datasets separately, by using linear models for microarray (LIMMA) approach through GEO2R online tool [22] with $|logFC| > 1.0$ and adjusted p-value < 0.05. The LIMMA approach calculates the p-value by using the modified t-statistics to test the significance of differential gene expressions between two conditions, and p-value is then adjusted by using the Benjamini–Hochberg procedure [23]. The p-value below a threshold 0.05 means that the expected proportion of false discoveries are less than 5%. Finally, we selected the common DEGs by using four DEGs sets derived from CC and EC based four publicly available microarray datasets.

2.6. Construction of Protein–Protein Interaction (PPI) Network for Identification of SGBs

The PPI network of common-DEGs was constructed through the STRING online database (https://string-db.org/) [24]. To improve the quality of PPI network, we used the Cytoscape software [25]. The Cytoscape plugin cytoHubba was used to select the common Hub Genes (cHubGs) or common Hub Proteins (cHubPs) from PPI network [25,26]. The PPI network provides several nodes and edges, which indicate proteins and their interactions, respectively. A node with the largest number of significant interactions/connections/edges with other nodes is considered as the top-ranked cHubGs. The cHubGs were selected by using four topological analyses (Degree [27], BottleNeck [28], Betweenness

[29], and Stress [30]) of the PPI network. Molecular Complex Detection (MCODE) (http://apps.cytoscape.org/apps/mcode) plugin of Cytoscape software was used to detect the most profound modules from the PPIs network. Highly interconnected portions were identified through MCODE clustering that assists the research in effective drug designing. For representing molecular complexes in the PPI network, MCODE was used by detecting the densely connected areas [31]. Then, we selected the shared genetic biomarkers (SGBs) that were shared by both cHubGs and MCODE clustering genes.

2.7. Regulatory Network Analysis of SGBs

To explore key transcriptional regulatory transcription factors (TFs) and post-transcriptional regulatory micro-RNAs (miRNAs) of KGs, we performed TFs–TGs and miRNAs–KGs interaction network analysis, respectively, by using the NetworkAnalyst web platform [32]. The TFs–KGs and miRNAs–KGs interaction networks were constructed by using the ENCODE (https://www.encodeproject.org/) [33] and RegNetwork repository [34] databases, respectively. The Cytoscape software [25] was used to improve the quality of networks.

2.8. Gene ontology and KEGG pathway enrichment analysis of SGBs

Gene ontology (GO) functional and Kyoto Encyclopedia of Genes and Genomes (KEGG) pathway enrichment/annotation/over-representation analysis [35,36] is a widely used approach to determine the significantly annotated/enriched/over-represented functions/classes/terms and pathways by the identified cDEGs/cHubGs. It is an important part for revealing the molecular mechanisms of actions and cellular roles of genes. The GO terms are categorized into Biological Process, Cellular Component, and Molecular Function [37]. We performed GO and KEGG enrichment analysis using Enrichr web tool (https://maayanlab.cloud/Enrichr/)[38]. The significance level was set to p-value < 0.05.

2.9. Survival Analysis

Gene Expression Profiling Interactive Analysis 2 (GEPIA2) is a new online database for profiling cancer and normal gene expression [39]. In this study, we analyzed the overall survival of CC and EC patients through the expressions of SGBs by using the CESC and UCEC dataset in the GEPIA2 database. We divided patients into two groups (the high and low expression groups) according to the median expression level of each SGB. The overall survival curves of each SGBs were calculated and plotted using the GEPIA2 database, with hazard ratio (HR) with 95% confidence intervals and log-rank P value.

2.10. Drug Repurposing by Molecular Docking Study

The molecular docking simulation study was adopted to perform the interaction among the target receptors and the drug molecules. Basically, this study interprets the potential drug components based on computational binding affinity. In this study, the docking analysis would be performed among the drug target key receptors biomolecule and 145 drug agents or small compounds (Table S2). The three dimensional (3D) structures of receptors were downloaded from Protein Data Bank (PDB) [40] and SWISS-MODEL [41], and the PubChem database [42] was used to retrieve the 3D structures of 175 meta-drug agents. The "Discovery Studio Visualizer" was used to visualize the 3D structures of protein interfaces [43]. PDB2PQR and H++ servers were utilized to assign the protonation state of target proteins [44,45]. All the missing hydrogen atoms were also appropriately added. The pKa for target proteins residues were investigated under the physical conditions of salinity = 0.15, internal dielectric = 10, pH = 7, and external dielectric = 80. The receptor proteins were preprocessed by removing water molecules and adding charges using AutoDock tools [46]. The drug agents were minimized energy through the Avogadro [47] and pre-processed by AutoDock tools [46], respectively. Subsequently, molecular docking between receptors and drug agents were performed to calculate their binding affinities (kcal/mol) by using AutoDock Vina [48]. The Discovery Studio Visualizer [40] and PyMol [49] was used to analyze the docked complexes for surface complexes, types and distances of non-covalent bonds For the details of the drug screening procedure see our previous published papers[50–53]. Then we validated our proposed drugs by molecular docking study with the top listed published receptor proteins associated with CC and EC infections that were obtained by the literature review. To select the top listed receptor proteins associated with CC and EC infections, we reviewed 66 recently published articles and selected the top listed 10 receptor proteins.

2.11. Molecular Dynamic (MD) Simulations

To explore the dynamic characteristics of the best protein-ligand complexes, MD simulations were performed using the YASARA software [54] and the AMBER14 force field [55]. MD simulations were performed on a variety of individual systems. The systems involved the best three hits, UBE2C_Sorafenib, AURKA_paclitaxel and TOP2A_Sunitinib complexes, complexes associated with our suggested receptors.

Before to simulation, A TIP3P [56] water model in a simulation cell was used to optimize and solve the hydrogen bonding network of protein-ligand complexes. Periodic limit conditions were retained with a solvent concentration of 0.997 gL-$^1$. The primary energy minimization procedure of each simulation system, involving 41697, 96252, and 55410 atoms for UBE2C_Sorafenib, TOP2A_Sunitinib, and AURKA_paclitaxel, respectively complexes were done via a simulated annealing technique employing the steepest gradient approach (5000 cycles). Under physiological settings (298 K, pH 7.4, 0.9% NaCl), every simulation was performed through the multiple time-step technique [57] with a time-step interval of 2.50 fs [58]. A 100 ns MD simulation was run at Berendsen thermostat [59] and constant pressure. For more details about MD simulation methodology see our previous papers [3]. The trajectories were noted for every 250 ps for additional analysis, and the preceding analysis was carried out using the default script of the YASARA [60] macro and the SciDAVis program(http://scidavis.sourceforge.net/). After that, all snapshots were subjected to the MM-PBSA (MM-Poisson–Boltzmann surface area) binding free energy calculation in YASARA software using the following formula [61].

$$\text{Binding free Energy} = E_{potReceptor} + E_{solvReceptor} + E_{potLigand} + E_{solvLigand} - E_{potComplex} - E_{solvComplex}$$

Here, the binding energy of MM-PBSA was computed using YASARA built-in macros via AMBER 14 as a force field, with larger positive energies indicating better binding [11,62].

3. Results

3.1. Identification of DEGs

The datasets GSE63514 and GSE9750, GSE17025 and GSE63678 were analyzed to identify DEGs between infections and control samples for CC and EC, and the DEGs in each dataset were presented using the volcano plots (Figure 2a–d), where blue and red dots represented the up-regulated and down-regulated genes, respectively. In GSE17025, a total of 5122 DEGs with 1856 up-regulated and 3266 down-regulated genes; in GSE63678, a total of 524 DEGs with 272 up-regulated and 252 down-regulated genes; in GSE63514, a total of 4091 DEGs with 2631 up-regulated and 1460 down-regulated genes; in GSE9750, a total of 2640 DEGs with 711 up-regulated and 1929 down-regulated genes were identified by GEO2R online tool with $|logFC| > 1.0$ and adjusted p-value $< 0.05$. Then, we found 62 up-regulated common-DEGs and 10 down-regulated common-DEGs for CC and EC patients (see Figure 2e; Table 1).

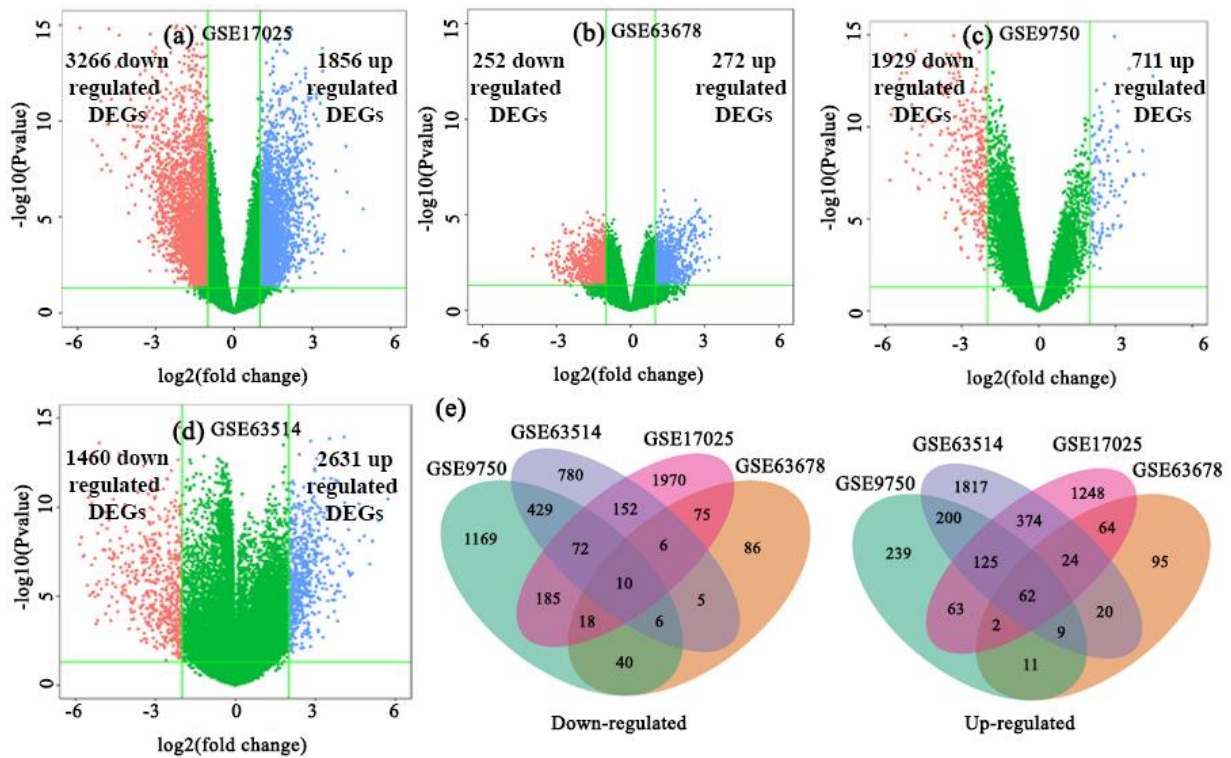

**Figure 2.** Screening of the sharing DEGs among GSE63514, GSE9750, GSE17025 and GSE63678 datasets. The volcano plots of DEGs in (**a**) GSE17025, (**b**) GSE63678, (**c**) GSE9750, and (**d**) GSE63514; blue dots and red dots represented the significantly up-regulated and down-regulated DEGs, respectively. (**e**) The sharing down-regulated and up-regulated DEGs between CC and EC visualized through a Venn diagram. Sixty-two genes were founded common up-regulated and ten genes were founded common down-regulated in CC patients.

Table 1. List of 72 DEGs, that were identified as upregulated or downregulated using the GSE6791, GSE27678, GSE63514, and GSE9750 microarray datasets.

| Type of Differentially Expressed Genes (DEGs) | DEGs |
|---|---|
| 62 Upregulated DEGs | NEK2, DTL, NUP210, FOXM1, CDC25A, AURKA, ISG15, TYMS, GTSE1, UBE2C, ECT2, SPAG5, KIF4A, KIF2C, KIF14, TK1, TACC3, TPX2, MELK, |
|  | MMP12, CDC6, CDC20, SLC35F6, CENPA, CEP55, TOP2A, CDK1, RRM2, BIRC5, FEN1, LMNB1, ZWINT, STAT1, CENPF, FANCA, SMC4, PRC1, CTSZ, CDCA3, ASPM, RAD51AP1, MKI67, SULF1, KIF20A, CENPN, ANP32E, HJURP, CCNB2, DLGAP5, CENPE, NUSAP1, KIF23, TTK, CDC7, UBE2S, NCAPG, CCNA2, HMMR, CKS2, CCNB1, CKAP2, E2F8 |
| 10 Downregulated DEGs | PDCD4, ZSCAN18, TBX3, PEG3, TGFBR3, GNAL, HNMT, C1orf21, PER1, MIR6883 |

3.2. PPI Network Analysis of common-DEGs for Identification of SGBs

The protein-protein interaction (PPI) network of common-DEGs was constructed using STRING database (Figure 3a), which contained 59 nodes and 1276 edges. We selected top-ranked 13 common hub genes (cHubGs) {AURKA, CCNB1, CCNB2, CDC20, CDC25A, CDC7, CDK1, ISG15, KIF2C, NUSAP1, TOP2A, UBE2C and UBE2S}, applying four topological measures in the PPI network. Then, using MCODE, clusters were selected from the PPI network. It was shown that the most significant cluster had 14 nodes and 91 edges. MCODE analysis demonstrated that the most significant cluster contained the nine hub genes {AURKA, CCNB1, CCNB2, CDC20, CDK1, KIF2C, NUSAP1, TOP2A, and UBE2C} (see Figure 3b). So, we considered these nine SGBs for further analysis.

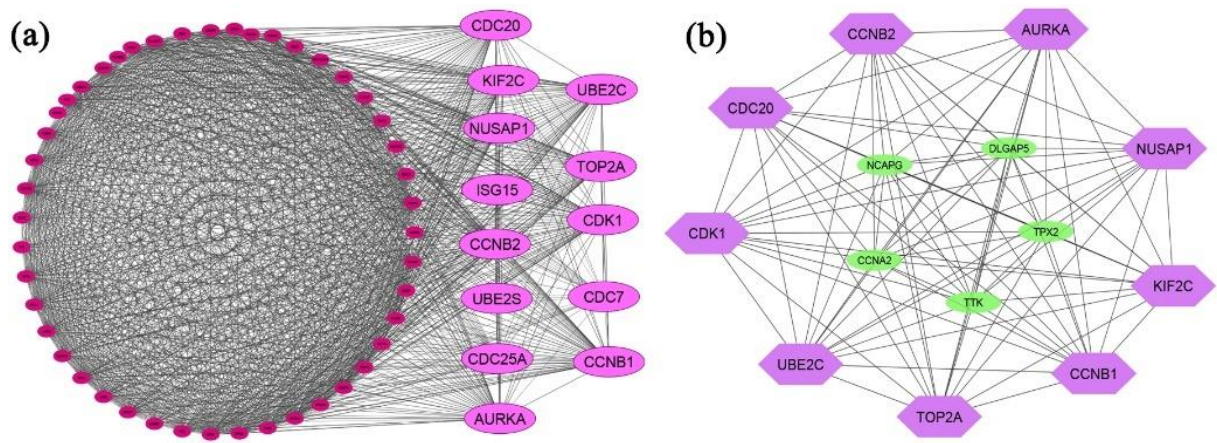

**Figure 3.** (**a**) Protein–protein interaction network for common differentially expressed genes of CC, and edges specify the interconnection in the middle of two genes. The analyzed network holds 59 nodes and 1276 edges. Surrounding nodes AURKA, CCNB1, CCNB2, CDC20, CDC25A, CDC7, CDK1, ISG15, KIF2C, NUSAP1, TOP2A, UBE2C and UBE2S) represented the hub genes. (**b**) Module analysis network obtained from MCODE analysis. Surrounding nodes (AURKA, CCNB1, CCNB2, CDC20, CDK1, KIF2C, NUSAP1, TOP2A, and UBE2C) were found to be common across 9 SGBs, so we considered these 9 genes as the key genes. The network represents highly interconnected regions of the PPI network. The network holds 14 nodes and 91 edges.

3.3. The Regulatory Network Analysis of SGBs

The network analysis of shared genetic biomarkers (SGBs) with transcription factors (TFs) detected top-ranked three significant TFs (FOXC1, NFIC and GATA2) as the key transcriptional regulatory factors for SGBs (see Figure 4a). We found FOXC1 as key TFs for six SGBs (AURKA, CCNB2, CDC20, CDK1, KIF2C, NUSAP1), NFIC for six

SGBs (AURKA, CCNB1, CDC20, CDK1, TOP2A, and UBE2C), and GATA2 for five SGBs (AURKA, CCNB2, CDK1, TOP2A, and UBE2C). Similarly, the network analysis of SGBs with miRNAs identified top-ranked seven significant miRNAs, denoted as hsa-mir-34a-5p, hsa-mir-107, hsa-mir-195-5p, hsa-mir-147a, hsa-mir-124-3p, hsa-mir-205-5p, and hsa-mir-129-2-3p, that were considered as the key post-transcriptional regulatory factors for all SGBs (see Figure 4b).

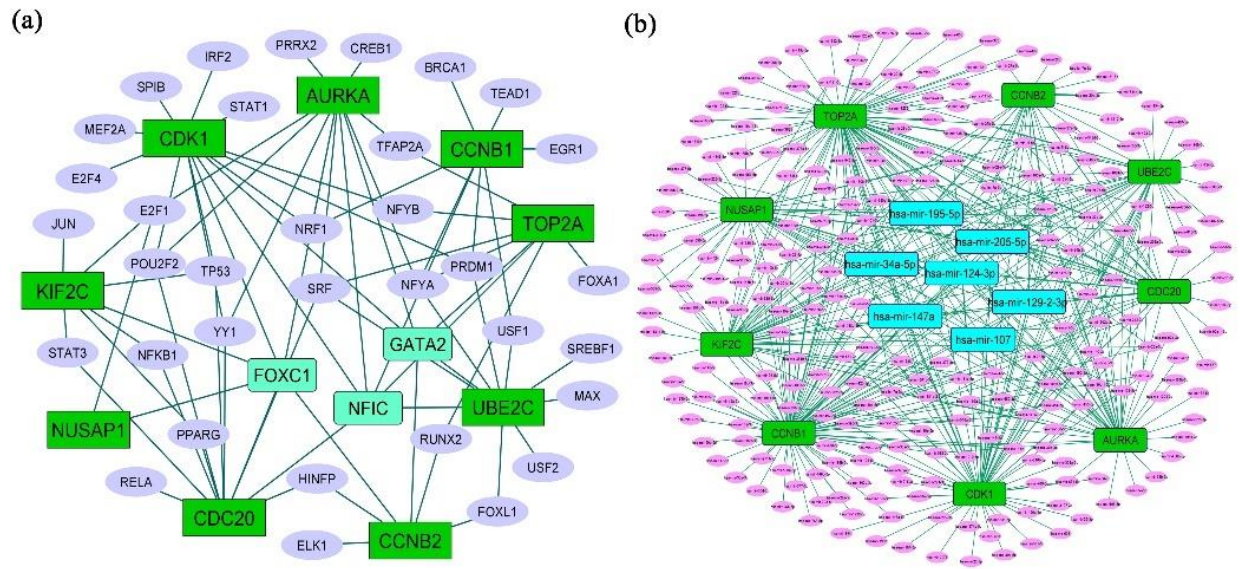

**Figure 4.** TFs-genes and miRNAs-genes interaction network with SGBs. a) Network for TFs-genes interaction with 9 SGBs. The highlighted dark green color nodes represented the SBGs and other nodes represent TF-genes and the highlighted turquoise color nodes represent the key TF-genes. The network consists of 46 nodes and 78 edges. b) The network presents the miRNA coregulatory network with 9 SGBs. The network consists of 262 nodes and 491 edges. The nodes in dark green color are represents the SGBs and other nodes represent miRNA and the highlighted cyan color nodes represent the key miRNA.

3.4. GO Functions and KEGG Pathway Enrichment Analysis of common-DEGs Highlighting SGBs

Table 2 displays the top five significantly enriched GO terms and KEGG pathways involving SGBs of CC and EC diseases, as supported by the literature review [9,10,63–71]. The top five GO terms for biological processes, including anaphase-promoting complex-dependent catabolic process, regulation of ubiquitin protein ligase activity, regulation of mitotic cell cycle phase transition, negative regulation of ubiquitin-protein ligase activity involved in the mitotic cell cycle, and mitotic cell cycle phase transition, were significantly enriched by the KGs sets {CDC20, CCNB1, UBE2C, CDK1, AURKA}, {CDC20, CCNB1, UBE2C, CDK1}, {CDC20, CCNB1, UBE2C, CDK1, AURKA}, {CDC20, CCNB1, UBE2C, CDK1}, and {CCNB2, CCNB1, UBE2C, CDK1, AURKA}, respectively. The molecular function (MF) GO terms, such as cyclin-dependent protein serine/threonine kinase activity, cyclin-dependent protein kinase activity, histone kinase activity, protein serine/threonine kinase activity, and protein kinase binding, were significantly enriched by the SGBs sets {CCNB2, CCNB1, CDK1}, {CCNB2, CCNB1, CDK1}, {Grande, 2020 #22}, {CCNB2, CCNB1, CDK1, AURKA}, and {TOP2A, CCNB1, AURKA}, respectively. The cellular component GO terms, including spindle, microtubule cytoskeleton, mitotic spindle, microtubule, and spindle microtubule, were significantly enriched by the SGBs sets {AURKA, CDC20, CCNB1, NUSAP1, CDK1, KIF2C}, {AURKA, CDC20, CCNB2, KIF2C}, {NUSAP1, CDK1, AURKA}, {AURKA, NUSAP1, CDK1, KIF2C}, and {NUSAP1, CDK1, AURKA}, respectively. We observed that KEGG pathways, including cell cycle, progesterone-mediated oocyte maturation, p53 signaling pathway, cellular senescence, and oocyte meiosis, were significantly enriched by the SGBs sets {CDC20, CCNB2, CCNB1, CDK1}, {CCNB2, CCNB1, CDK1, AURKA}, {CCNB2, CCNB1, CDK1}, {CCNB2, CCNB1, CDK1}, and {CDC20, CCNB2, CCNB1, CDK1, AURKA}, respectively.

**Table 2.** The top five significantly (p-value < 0.001) enriched GO functions and KEGG pathways by cDEGs involving KGs with CC diseases.

| | GO ID | GO Term | *p*-Value | Associated KGs |
|---|---|---|---|---|
| **Biological Process (BPs)** | GO:0031145 | Anaphase-promoting complex-dependent catabolic process | 1.05E-10 | CDC20, CCNB1, UBE2C, CDK1, AURKA |
| | GO:1904666 | Regulation of ubiquitin protein ligase activity | 1.11E-09 | CDC20, CCNB1, UBE2C, CDK1 |
| | GO:1901990 | Regulation of mitotic cell cycle phase transition | 7.63E-09 | CDC20, CCNB1, UBE2C, CDK1, AURKA |
| | GO:0051436 | Negative regulation of ubiquitin-protein ligase activity involved in mitotic cell cycle | 1.81E-08 | CDC20, CCNB1, UBE2C, CDK1 |
| | GO:0044772 | Mitotic cell cycle phase transition | 1.91E-08 | CCNB2, CCNB1, UBE2C, CDK1, AURKA |
| **Molecular Function (MFs)** | GO:0004693 | cyclin-dependent protein serine/threonine kinase activity | 3.41E-07 | CCNB2, CCNB1, CDK1 |
| | GO:0097472 | cyclin-dependent protein kinase activity | 3.74E-07 | CCNB2, CCNB1, CDK1 |
| | GO:0035173 | histone kinase activity | 6.47E-06 | CDK1, AURKA |
| | GO:0004674 | protein serine/threonine kinase activity | 1.32E-05 | CCNB2, CCNB1, CDK1, AURKA |
| | GO:0019901 | protein kinase binding | 0.001132505 | TOP2A, CCNB1, AURKA |
| **Cellular Component** | GO:0005819 | spindle | 7.56E-23 | AURKA, CDC20, CCNB1, NUSAP1, CDK1, KIF2C |
| | GO:0015630 | microtubule cytoskeleton | 1.63E-15 | AURKA, CDC20, CCNB2, KIF2C |
| | GO:0072686 | mitotic spindle | 8.94E-15 | NUSAP1, CDK1, AURKA |
| | GO:0005874 | microtubule | 5.34E-13 | AURKA, NUSAP1, CDK1, KIF2C |
| | GO:0005876 | spindle microtubule | 4.42E-10 | NUSAP1, CDK1, AURKA |
| | **Pathways** | | *p*-Value | **Associated cHubGs** |
| **KEGG Pathway** | | Cell cycle | 6.18E-10 | CDC20, CCNB2, CCNB1, CDK1 |
| | | Progesterone-mediated oocyte maturation | 1.51E-06 | CCNB2, CCNB1, CDK1, AURKA |
| | | p53 signaling pathway | 6.09E-06 | CCNB2, CCNB1, CDK1 |
| | | Cellular senescence | 2.41E-05 | CCNB2, CCNB1, CDK1 |
| | | Oocyte meiosis | 8.81E-05 | CDC20, CCNB2, CCNB1, CDK1, AURKA |

3.5. Survival Analysis with SGBs

The GEPIA2 database was used to analyze the prognostic value of these 9 SGBs in CC and EC patients using the TCGA-CESC and TCGA-UCEC cohort. The survival analysis results showed that only three SGBs AURKA (log-rank p=0.051), UBE2C (log-rank p=0.076), and TOP2A (log-rank p=0.029) were significantly negatively correlated with the prognosis of CC and EC patients (Figure 5a), while the others genes showed no statistical significance in the overall survival of CC and EC patients (supplementary Figure-S1). On the other hand, the expression level of these three genes among CC and EC patients were different from the overall survival time of CC and EC patients. Compared with normal tissues, these three genes expression were significantly increased in CC and EC (Figure 5b)

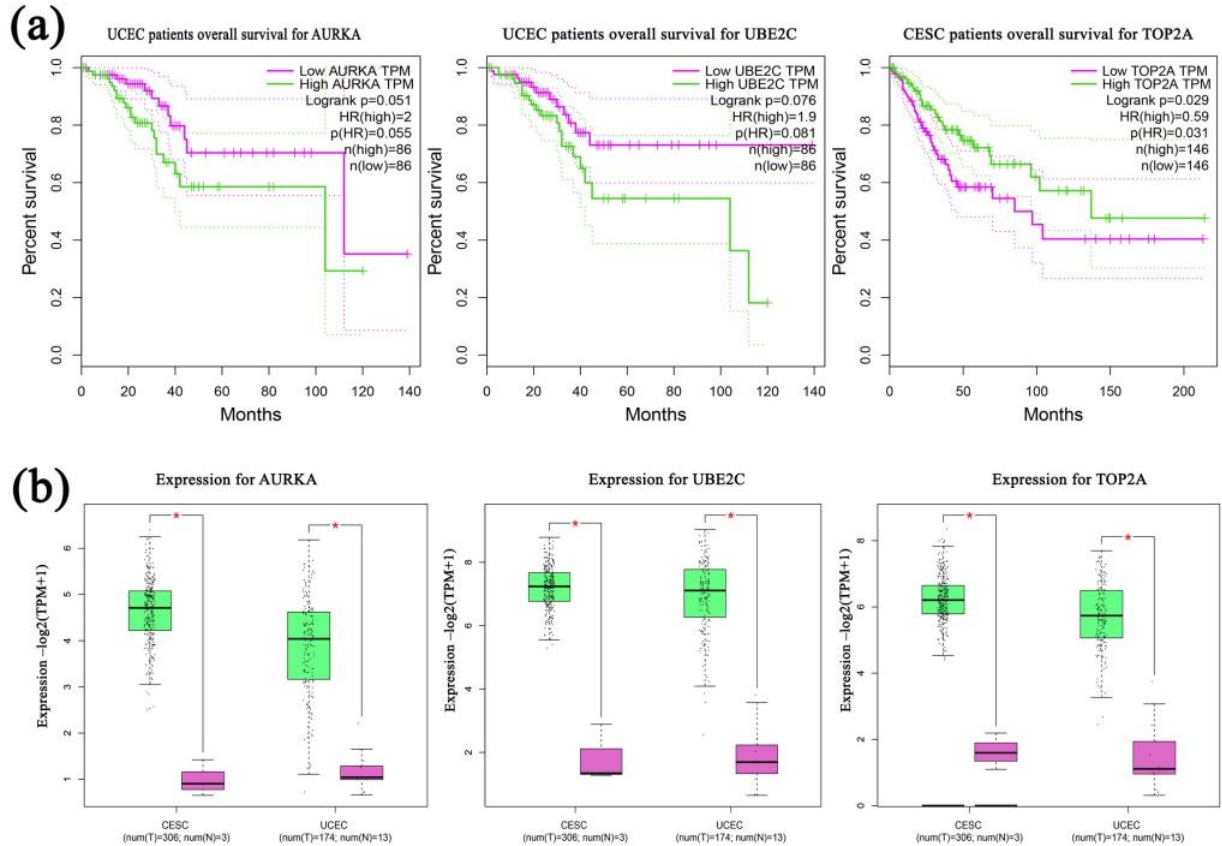

**Figure 5.** The overall survival curves of 3 significant SGBs in CC and EC patients. The green curve is the high expression group and the magenta curve is the low-expression group. Data are presented as the hazard ratio with a 92% confidence interval and log rank p-value < 0.08.

3.6. Drug Repurposing by Molecular Docking

We selected 9 potential hub receptor proteins and 145 drug agents to explore candidate drugs by molecular docking study for the treatment against CC and EC patients (see Table S2). We downloaded 3D structure of three drug target proteins (AURKA, UBE2C, TOP2A, CCNB1, CDK1, CDC20, and KIF2C) from Protein Data Bank (PDB) [40] with source codes 1mq4, 1i7k, 1zxm, 2b9r, 4y72, 4ggc, and 2heh, respectively as well as the 3D structure of remains two proteins (CCNB2 and NUSAP1) were downloaded from SWISS-MODEL [41] using UniProt [72] ID of O95067 and Q9BXS6, respectively. The 3D structures of 145 drugs (see Table S2) were downloaded from PubChem database [42] as mentioned previously.

On the other hand, we reviewed 66 published articles associated with CC and EC infections that provided transcriptome-guided hub/studied biomarkers for cross-validation of the proposed key genes and the candidate drug agents. We considered top-ranked 10 Key genes (KGs) (CCNB1, CDC45, MCM2, TOP2A, CDK1, AURKA, BIRC5, CDC20, CDCA8, and CENPA) that were published in at least three articles (see Table S1). Five (CDC20, TOP2A, CDK1, AURKA, and CCNB1) of the ten reported hub/studied genes were found to be similar to our suggested nine SGBs. So, we downloaded the 3D structure of others receptors (BIRC5, MCM2, CDCA8, CENPA, and CDC45) from Protein Data Bank (PDB) [40] with source codes 1e31, 4uuz, 2kdd, 3nqj, and 5dgo, respectively. Then, the molecular docking analysis was performed among total 14 target receptors (proposed and published) and 145 meta-drug agents to calculate the binding affinities (kcal/mol) for each pair of drug targets and agents. Then, for the selection of few numbers of candidate drug agents, we ordered the target proteins according to the row sums of the binding affinity matrix A=(Aij) and the column sums of the drug agents (see Figure 6). Therefore, we selected top-ranked ten candidate drug agents (Sorafenib, Paclitaxel, Sunitinib, Vincristine, Vinorelbine, Rapamycin, Ridaforolimus, Everolimus, Docetaxel, and Temsirolimus) as candidate drug agents with binding affinity scores −7.6 kcal/mol <= against the 14 target proteins.

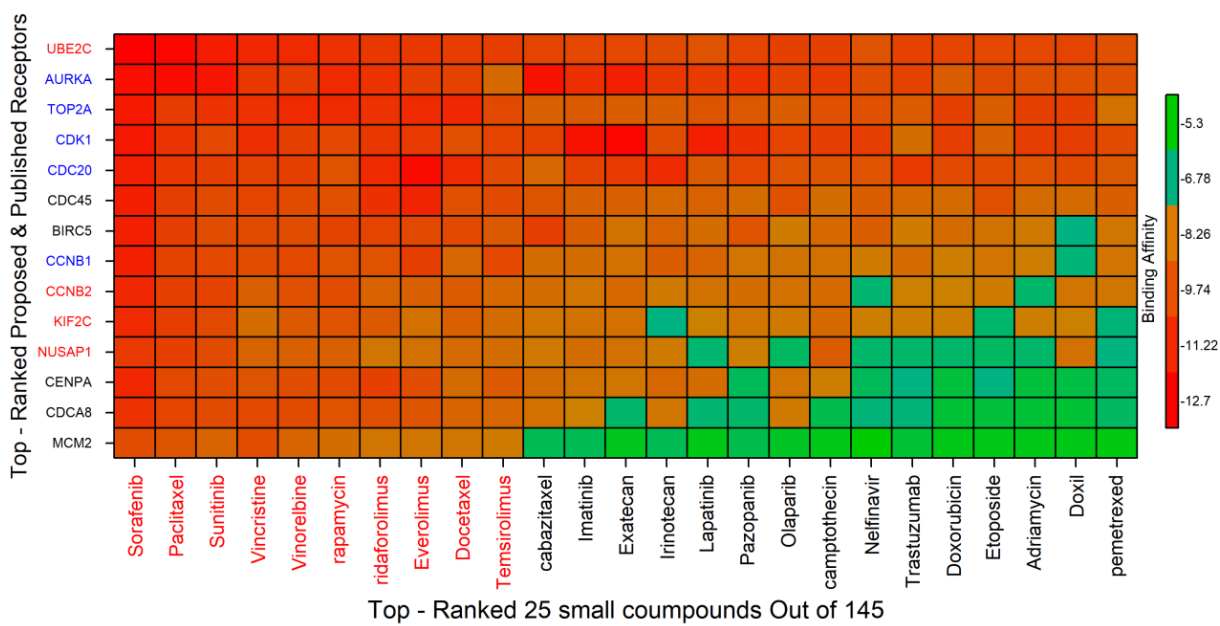

**Figure 6.** Molecular docking study scores calculated through Autodock Vina, where, the red and green colors indicated the strong and weak docking scores between target proteins and small molecules. In Y-axis, blue color

proteins indicated common receptors (published and proposed), whereas red and black color proteins indicated only proposed and published receptors, respectively.

The docked complexes of the top three virtual hits from the docking analysis were further considered for protein–ligand interaction profiling. As shown in Figure 7a, UBE2C_Sorafenib complex showed three hydrogen bonds with SER134, SER137, SER114 residues. Although the ligand formed major hydrophobic interactions with LEU133 and LEU115 residues. Residue ASP116 and LEU133 showed additional pi-anion and halogen interactions with the ligand, respectively. On the other hand, AURKA_Paclitaxel (Figure 7b) complex showed only one hydrogen bonds with LYS141 residue, and three hydrophobic interactions with LEU139, LEU263, TYR219 residues. Residue LYS162, GLU181, ASP274 showed additional attractive charge with the ligand. In the case of the TOP2A_Sunitinib complex, Sunitinib showed one hydrogen bond with SER320 residue. Sunitinib also formed hydrophobic interactions with LE311 and ALA318 residues. Furthermore, sunitinib also formed halogen interactions with SER312 residue (see Figure 7c).

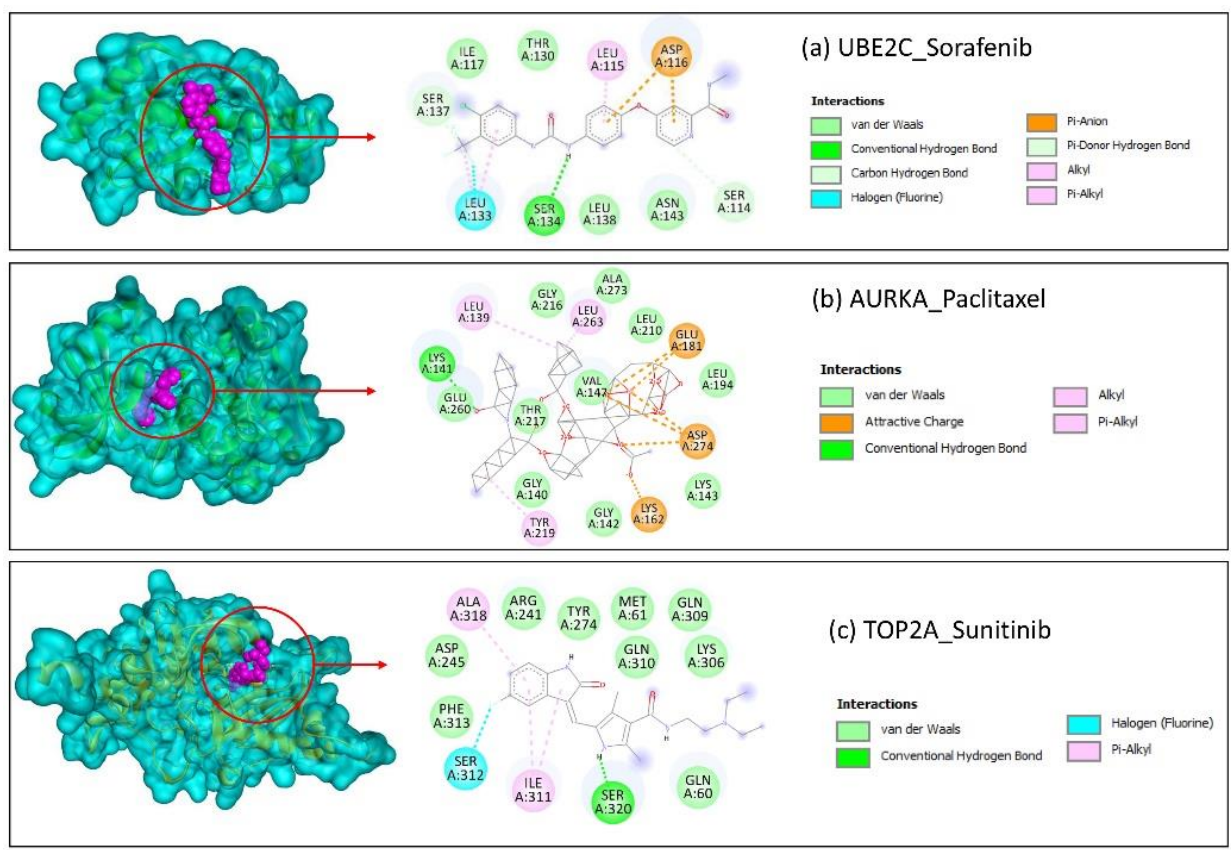

**Figure 7.** The 3-dimension view of strong binding interactions between targets and drugs is shown in the 1st column and their two-dimensional chemical interactions showed in middle. The last column showed their binding types with potential targets. Complexes: (a) UBE2C_Sorafenib, (b) AURKA_Paclitaxel, and (c) TOP2A_Sunitinib

3.7. MD Simulation Study

Among the proposed candidate drugs, Sorafenib, Paclitaxel and Sunitinib were the top-ranked three candidate drugs (Figure 6). Therefore, these top three drug agents were selected for their stability analysis through 100 ns MD-based MM-PBSA simulations.

From Figure 8, we observed that all the three systems were significantly stable between the variations of moving and initial drug–target complexes. We calculated their RMSD (root mean square deviation). Figure 8a represented the RMSD corresponding to the proposed receptors (UBE2C, AURKA and TOP2A). All the systems projected the RMSD around 1 Å to 2.5 Å, except TOP2A complex, which shows the RMSD around 2 Å to 3.7 Å. The average RMSD for UBE2C, AURKA and TOP2A complexes were 1.59 Å, 2.11 Å, and 2.80 Å, respectively. UBE2C complex showed slight fluctuation in around 20,000 ps to 28,000 ps and was stabilized in the remaining simulation. As can be seen from the plot, AURKA showed a more rigid conformation than the other proteins, also achieved equilibrium at 3 ns, and remained stable afterward. In contrast, TOP2A showed a dramatic increase in flexibility, with RMSD values rising gradually from 2 Å to 3.5 Å over time. Here, we calculated the MM-PBSA binding energy for three drug agents as mentioned previously, Figure 8b represented the binding energy with the top-ranked three proposed potential biomarkers (UBE2C, AURKA and TOP2A). On an average, UBE2C, AURKA and TOP2A complexes produced binding energies −192.65 kJ/mol, −26.36 kJ/mol, and 41.56 kJ/mol, respectively.

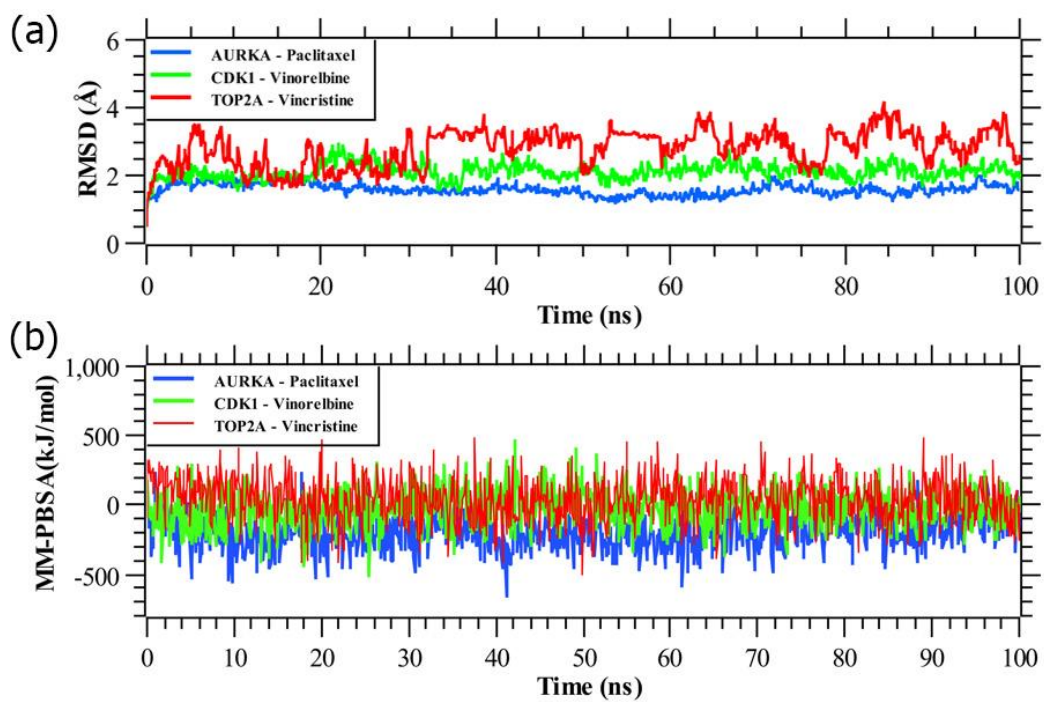

**Figure 8.** (**a**) Time evolution of root mean square deviations (RMSDs) of backbone atoms (C, Cα, and N) for protein for each docked complex. (**b**) Binding free energy (in kJ mol$^{-1}$) of each snapshot was calculated by molecular mechanics Poisson–Boltzmann surface area (MM-PBSA) analysis, representing the change in binding stability of each complex during simulations; positive values indicate better binding. Complexes: green UBE2C, blue AURKA, and red TOP2A.

## 4. Discussion

Cancer research has advanced significantly in recent years, notably in identifying etiological causes. Although several studies have examined CC and EC individually, few have focused on these two types of cancer in combination. The aim of the present study was to perform a computational analysis to discover potential SGBs and candidate drugs to improve the 5-year survival rate and reduce the mortality rate of CC and EC patients, and to make a preliminary assessment of the associations between these two types of cancer.

By analyzing four publicly available microarray datasets, 12377 DEGs were identified, including 6731 in CC and 5646 in EC. To investigate the genetic influence on CC and EC infections, we identified seventy-two shared common-DEGs. Among them, we detected nine SGBs, highlighting their functions, pathways, regulatory factors, and candidate drugs. The literature review suggested that the Ubiquitin-conjugating enzyme E2C (UBE2C) was highly expressed in endometrial and cervical cancer [73]. AURKA is frequently amplified in several tumors, including CC, EC, breast, pancreatic, colorectal, gastric, and ovarian carcinomas, and in some cases, is associated with poor prognosis [3]. TOP2A is closely related to the pathogenesis of CC and EC [74]. CCNB1 plays vital roles in the progression of CC and EC through different signaling pathways [50,75] CDK1 is considered a biomarker for the improved diagnosis of CC and EC [3,76] CCNB2 has been identified as one of the most important therapeutic drug targets for CC and EC, which are among the most life-threatening and devastating tumors [75]. CDC20 may be a promising novel therapeutic target for CC and EC treatment [75,77] KIF2C is strongly associated with the survival rate of CC and EC [78]. NUSAP1 expression is upregulated in CC and contributes to metastasis by enhancing cancer stem cell traits and epithelial-to-mesenchymal transition (EMT) [79]. We also examined the overall survival of CC and EC patients with SGBs and observed a significant difference between the two survival probability curves, showing that the suggested SGBs had a stronger prognostic capacity for CC and EC diagnosis. The TFs versus SGBs interaction network analysis exposed 3 key TFs-proteins (FOXC1, NFIC, and GATA2) as the transcriptional regulatory factors of Hub-DEGs (Figure 4a). These 3 TFs proteins were further used as drug target receptors. We also identified seven post-transcriptional (miRNAs) regulatory factors of SGBs (Figure 4b). Previous studies also supported that FOXC1 contributed to the development of CC and EC by increasing the growth and motility of cancer cells [80,81]; NFIC is associated with poor overall survival in CC and EC [82,83]; GATA2 is associated with the recurrence of EC and deregulated with poor survival outcomes [84].

To investigate the common pathogenetic processes of SGBs, we selected the top five common GO terms for each of BPs, MFs, and cellular components, and KEGG pathways that were significantly associated with CC and EC infections through the common-DEGs, including SGBs. Among them, the association of the top five common BPs (anaphase-promoting complex-dependent catabolic process, regulation of ubiquitin protein ligase activity, regulation of mitotic cell cycle phase transition, negative regulation of ubiquitin-protein ligase activity involved in the mitotic cell cycle, and mitotic cell cycle phase transition) with CC and EC was supported by some other individual studies [85–87]. The top five common MFs (cyclin-dependent protein serine/threonine kinase activity, cyclin-dependent protein kinase activity, histone kinase activity, protein serine/threonine kinase activity, and protein kinase binding) that were significantly associated with CC and EC disease also received support from some individual studies [85,88]. Similarly, the association of the top four cellular components (spindle, microtubule cytoskeleton, mitotic spindle, microtubule, and spindle microtubule) with CC and EC disease was supported by the literature review [87,88]. We selected the top five significantly enriched common KEGG pathways (cell cycle, progesterone-mediated oocyte maturation, p53 signaling pathway, cellular senescence, and oocyte meiosis) that were also reported by some other studies [85,87,88].

To explore effective candidate drugs for the treatment against CC and EC disease, we considered top-ranked proposed receptors as the drug target receptors and performed their docking analysis with 145 meta-drug agents (see Table S2). For cross-validation, we considered top-ranked ten published hub proteins, where five proteins were common with our proposed SGBs (see Table S1). So, molecular docking was carried out between total m = 14 receptors (proposed and published) and n = 145 meta-drug agents to calculate the binding affinity scores (kcal/mol) for each pair of receptors and drug agents.  Then we selected top-ranked ten drugs (Sorafenib, Paclitaxel, Sunitinib, Vincristine, Vinorelbine, Rapamycin, Ridaforolimus, Everolimus, Docetaxel, and Temsirolimus) as the potential drugs for EC and CC infections based on their strong binding affinity scores (kCal/mol) with all the target receptors (see Figure 6). Some other independent studies also recommended our suggested drugs including Sorafenib [89–92], Paclitaxel [93–95], Sunitinib [89–91]**,** Docetaxel [89,90], Temsirolimus [90,91], Vincristine [3,93,96], Everolimus [91,97] Vinorelbine [3,96,98], Rapamycin [92], and Ridaforolimus[92] for the treatment of EC and CC infections. Finally, we examined the stability of top-ranked three drugs (Sorafenib, Paclitaxel, Sunitinib) by using 100 ns MD-based MM-PBSA simulations for three top-ranked proposed receptors (UBE2C, AURKA, TOP2A) and observed their stable performance according to the laws of physics [99,100]. Therefore, the proposed candidate drugs might play vital role in the treatment of CC and EC infections.

## 5. Conclusions

The pathogenesis of CC and EC is complicated. The present study utilized various well-established bioinformatics tools to reveal SGBs, highlighting their regulatory factors and dysregulated molecular functions and pathways responsible for CC and EC progression. By performing a computational analysis, we revealed a library of DEGs in CC and EC and identified nine SGBs. Five of these SGBs were common between our selected nine SGBs and the top-ranked ten studied/hub genes proposed by others, which indicates that our proposed SGBs have received more support from the literature review compared to other individual studies. Finally, we offered potential candidate drugs, such as Sorafenib, Paclitaxel, and Sunitinib, and examined their stability performance using 100 ns MD-based MM-PBSA simulations for the top-ranked three proposed proteins (UBE2C, AURKA, and TOP2A), also detecting their stable performance. Hence, the selected molecular biomarkers and repurposing candidate drugs presented in this research have merit for the diagnosis and therapy of CC and EC diseases.

**Supplementary Materials:** The following supporting information can be downloaded at: www.mdpi.com/xxx/s1, **Table S1.** Different hub/studied genes list for cervical cancer (CC) and endometrial carcinoma (EC) infection published by different papers in different reputed journals, **Table S2:** Meta-drug agents suggested by different papers in different reputed journals for the treatment against CC and EC infections.

**Author Contributions:**

Md. Selim Reza: Conceptualization, methodology, software, validation, formal analysis, writing—original draft preparation, review and editing.

Mst. Ayesha Siddika: validation, investigation, review and editing.

Md. Tofazzal Hossain: validation, investigation, review and editing.

Md. Ashad Alam: validation, investigation, review and editing.

Md. Nurul Haque Mollah2: validation, investigation, supervision, review and editing.

All authors have read and agreed to the published version of the manuscript.

**Funding:** No fund received.

**Institutional Review Board Statement:** Not applicable.

**Informed Consent Statement:** Not applicable

**Data Availability Statement:**

The datasets analyzed in this study were downloaded with accession numbers GSE63514, GSE9750, GSE17025 and GSE63678 from the NCBI- GEO database (https://www.ncbi.nlm.nih.gov/gds/). The dataset GSE63514 was obtained from the GPL570 platform ((HG-U133_Plus_2) Affymetrix Human Genome U133 Plus 2.0 Array). It included 28 CC tissue and 24 normal tissue, while the GSE9750 data based on the GPL96 platform ([HG-U133A] Affymetrix Human Genome U133A Array) included 33 CC tissue and 21 normal tissue. The GSE17025 data based on the GPL570 platform ([HG-U133_Plus_2] Affymetrix Human Genome U133 Plus 2.0 Array) included 79 EC tissues and 12 normal tissues. Moreover, the EC patient's dataset GSE63678 was generated using Affymetrix Human Genome U133A 2.0 Array technology with 5 normal tissues and 7 EC tissues; analysis was performed using platform GPL571. A part of the relevant data is available within the paper and its supporting information files.

**Conflicts of Interest:** The authors declare no conflicts of interest.

**Abbreviations:**-

| | |
|---|---|
| EC | Endometrial Cancer |
| UCEC | Uterine Corpus Endometrial Carcinoma |
| CC | Cervical Cancer |
| CSCC | Cervical Squamous Cell Carcinoma |
| HPV | Human Papillomavirus |
| LIMMA | Linear Models for Microarray Data |
| PPI | Protein–Protein Interaction |
| ENCODE | Encyclopedia Of DNA Elements |
| MCODE | Molecular Complex Detection |
| DEGs | Differentially Expressed Genes |
| Common-DEGs | Common Differentially Expressed Genes |
| SGBs | Shared Genetic Biomarkers |
| KPs | Key Proteins |
| DR | Drug Repurposing |
| GO | Gene Ontology |
| BPs | Biological Process |
| MFs | Molecular Functions |
| CCs | Cellular Components |
| KEGG | Kyoto Encyclopedia of Genes and Genomes |
| TFs | Transcription Factors |
| miRNAs | Micro-RNAs |
| MD | Molecular Dynamic |
| MM-PBSA | Molecular Mechanics Poisson–Boltzmann Surface Area |
| RMSD | Root Mean Square Deviation |
| 3D | Three-Dimensional |
| PDB | Protein Data Bank |
| YASARA | Yet Another Scientific Artificial Reality Application |